\documentclass[a4paper,12pt]{article}
\usepackage{epsfig}

\def\lsim{\raise0.3ex\hbox{$\;<$\kern-0.75em\raise-1.1ex\hbox{$\sim\;$}}}
\def\gsim{\raise0.3ex\hbox{$\;>$\kern-0.75em\raise-1.1ex\hbox{$\sim\;$}}}

\def\bmat{\left(\begin{array}}
\def\emat{\end{array}\right)}
\def    \be            {\begin{equation}}
\def    \ee            {\end{equation}}
\def    \bea           {\begin{eqnarray}}
\def    \eea           {\end{eqnarray}}

\begin{document}


%
%
%
%
%

\rightline{FTUAM 02/11}
\rightline{IFT-UAM/CSIC-02-11}
\rightline{HIP-2002-17/TH}
\rightline{hep-ph/0204271}
\rightline{April 2002}   

\renewcommand{\thefootnote}{\fnsymbol{footnote}}
\setcounter{footnote}{0}
\vspace{0.0cm}
\begin{center}
{\large{\bf Experimental Constraints on the Neutralino-Nucleon 
Cross Section\footnote{Talk given at Corfu Summer Institute 
on Elementary Particle Physics, August 31-September 20, 2001.}
\\[5mm]
}}
\vspace{0.5cm}
\mbox{
\sc{
D.G. Cerde\~no$^{1}$,
E. Gabrielli$^{2}$,
C. Mu\~noz$^{1}$
}
}
\vspace{0.5cm}

{\small
{\it $^1$ Departamento de F\'{\i}sica
Te\'orica C-XI and Instituto de F\'{\i}sica
Te\'orica C-XVI,\\ 
Universidad Aut\'onoma de Madrid,
Cantoblanco, 28049 Madrid, Spain. \\
\vspace*{2mm}
\it $^2$ Helsinki Institute of Physics,
P.O. Box 64, FIN-00014 Helsinki, Finland.
} 
}
\end{center}


\renewcommand{\thefootnote}{\alph{footnote}}
\setcounter{footnote}{0}

\vspace{0.2cm}

{\small 
In the light of recent experimental results for the direct detection
of dark matter,
we analyze in the framework of SUGRA the
value of the neutralino-nucleon cross section.
We study how this value is modified when the usual 
assumptions of universal soft terms and GUT scale are relaxed.
In particular we consider scenarios with non-universal
scalar and gaugino masses and scenarios 
with intermediate unification scale.
We also study superstring constructions with D-branes,
where a combination of the above two scenarios arises naturally.
In the analysis we take into account 
the most recent experimental constraints, such as
the lower bound on the Higgs mass,
the $b\to s\gamma$ branching ratio, and the
muon $g-2$.
}

\section{Introduction}

One of the most interesting candidates for dark matter is a 
long-lived or stable weakly-interacting massive 
particle (WIMP). WIMPs can remain from the earliest moments of the Universe in 
sufficient number to account for a significant fraction of relic density. 
These particles would form not only a background density in the Universe, but
also would cluster gravitationally with ordinary stars in the galactic halos.
This raises the hope of detecting relic 
WIMPs directly, by observing their elastic scattering on  
target nuclei through nuclear recoils \cite{contemporary}.
In fact, one of the current experiments, the DAMA collaboration,
reported recently \cite{experimento1}
data favoring the existence of a 
WIMP signal.
It was claimed that (at 4$\sigma$ C.L.) 
the preferred range of the WIMP-nucleon cross section
is 
$10^{-6}$-$10^{-5}$ pb 
for a WIMP mass between 30 and 200 GeV.
Unlike this spectacular result, other collaborations, 
CDMS \cite{experimento2}, EDELWEISS \cite{edelweiss}, 
and IGEX \cite{igex}, 
claim to have excluded regions of the DAMA 
parameter space.


In any case, due to these and other projected experiments, 
it seems very plausible that the dark matter 
will be found in the near future. 
In this situation, and assuming that the dark matter 
is a WIMP, it is natural to wonder how big 
the cross section for its direct detection can be.
The answer to this 
question depends on the particular WIMP considered.
The leading candidate in this class is the lightest 
neutralino
\cite{contemporary}, 
a particle 
predicted by the supersymmetric (SUSY) extension of the standard model (SM).

In the simplest SUSY extension, the so-called
minimal supersymmetric standard model (MSSM),
there are four neutralinos, 
$\tilde{\chi}^0_i~(i=1,2,3,4)$, 
since they
are the physical 
superpositions of the bino ($\tilde{B}^0$), of the neutral 
wino 
($\tilde{W}_3^0$), 
and of the
neutral Higgsinos ($\tilde{H}^0_u$, $\tilde{H}_d^0$). 
In most of the parameter space of
the MSSM,
the lightest neutralino, $\tilde{\chi}^0_1$,
is the lightest supersymmetric particle (LSP). When R-parity is
imposed this implies that $\tilde{\chi}^0_1$ is absolutely stable,
and therefore a dark matter candidate.

In this paper we will analyze this SUSY scenario
in the framework of supergravity (SUGRA), and in particular
several constructions proposed recently in the 
literature (for a review see ref.~\cite{darkcairo}), where the 
$\tilde{\chi}^0_1$-nucleon
cross section can be enhanced, and might be of the order of $10^{-6}$ pb,
i.e., where current dark matter detectors are 
sensitive.
First, in Section~2, we will briefly introduce the SUGRA framework.
In Section~3 we will describe the most recent experimental 
constraints which can affect the computation
of the cross section. Then, in the rest of the sections, we 
will re-evaluate previous computations, taking into account
these constraints. In particular, 
in Section~4
we will review the value of the cross section when
universal soft SUSY-breaking 
terms are assumed,
and how
this value is modified when this assumption is 
relaxed 
\cite{Bottino,Arnowitt,Nath2,darkcairo,Arnowitt3}.
In Section~5 we will consider the case of an intermediate unification 
scale \cite{muas,bailin}.
Finally, in Section~6, we will study superstring scenarios with 
D-branes \cite{nosotros}.
 


\section{SUGRA framework}

Working in the framework of SUGRA one makes several 
assumptions. In particular, the soft parameters,
i.e., gaugino masses,
scalar masses, and 
trilinear couplings, are generated once SUSY is broken through
gravitational interactions.
They are denoted at the grand unification 
scale, $M_{GUT} \approx 2\times 10^{16}$ GeV,
by $M_{a}$,
$m_{\alpha}$, and $A_{\alpha\beta\gamma}$ 
respectively. 
Likewise, 
radiative electroweak symmetry breaking is imposed, i.e., 
the Higgsino mass parameter $\mu$ 
is determined by the minimization of the Higgs effective 
potential. This implies 
\begin{equation}
\mu^2 = \frac{m_{H_d}^2 - m_{H_u}^2 \tan^2 \beta}{\tan^2 \beta -1 } - 
\frac{1}{2} M_Z^2\ ,
\label{electroweak}
\end{equation} 
where
$\tan\beta= \langle H_u^0\rangle/\langle H_d^0\rangle$ 
is the ratio of Higgs vacuum expectation values.

The soft SUSY-breaking terms will have in general a
non-universal structure in the framework of SUGRA \cite{dilaton}. 
For the case of the observable scalar masses,
this is due to the non-universal couplings
in the K\"ahler potential
between the hidden sector fields breaking SUSY and the
observable sector fields.
For the case of the gaugino masses, this is due to the
non-universality of the gauge kinetic functions associated to the
different gauge groups.
Explicit string constructions, whose low-energy limit is SUGRA,
exhibit these properties \cite{dilaton}.

With these assumptions, the SUGRA framework still allows a 
large number of free 
parameters.
In particular, it contains more than 100 parameters.
In order to have predictive power one usually assumes in addition
that the above soft parameters are real and universal at $M_{GUT}$.
This is the so-called minimal supergravity (mSUGRA) scenario,
where there are only four free parameters: 
$m$, $M$, $A$, and $\tan \beta$. In addition, the
sign of $\mu$ remains also undetermined.  
It is worth noticing that explicit string constructions
with these characteristics 
can also be found \cite{dilaton}.


\section{
Experimental constraints}

Here we list the most recent experimental 
results which might be relevant when computing
the  $\tilde{\chi}^0_1$-nucleon
cross section in the context of SUGRA. In particular,
they may
produce important constraints in the
parameter space, forbidding for example points with a 
large cross section.

\begin{itemize}

\item Higgs mass

Whereas in the context of the SM,
the negative direct search for the Higgs at the LEP2 collider implies
a precise lower bound on its mass of 114.1 GeV,
the situation in SUSY scenarios is more involved.
In particular, in the framework of mSUGRA, 
one obtains \cite{Heinemeyer} for the lightest CP-even Higgs mass
$m_h \gsim 114.1$ GeV when $\tan\beta \lsim 50$,
and $m_h \gsim 91$ GeV when $\tan\beta$ is larger.
Let us remark that $\tan\beta$ is constrained to be 
smaller than 60 since otherwise the condition of
radiative electroweak symmetry breaking cannot be fulfilled.

On the other hand, when the mSUGRA framework is relaxed
the above SUSY bounds must be modified.
In particular, for the benchmark scenarios studied in 
ref.~\cite{experimentales} one obtains
$m_h \gsim 114.1$ GeV when $\tan\beta \lsim 8$,
and $m_h \gsim 91$ GeV when $\tan\beta$ is larger.

Let us finally remark that
in our computation below we evaluate $m_h$ using the 
program {\it FeynHiggsFast},
a simplified version of the updated program {\it FeynHiggs} \cite{FeynHiggs}
which contains the complete one-loop and dominant two-loop corrections.
The value of $m_h$ obtained with {\it FeynHiggsFast} is
approximately 1 GeV below the one obtained using {\it FeynHiggs}.
In addition, we should also keep in mind that 
the value of $m_h$ obtained with {\it FeynHiggs}
has an
uncertainty of about 3 GeV, due e.g. to higher-order corrections.

\item Top mass

Needless to say we use as input for the
top mass throughout this paper 
the central experimental value $m_t(pole)=175$ GeV.
However, let us remark that a modification in this mass by
$\pm 1$ GeV implies, basically, a modification also of $\pm 1$ GeV
in $m_h$.

\item SUSY spectrum

We impose
the present experimental lower
bounds on SUSY masses coming from LEP and Tevatron. In particular,
using the low-energy relation from mSUGRA,
$M_1=\frac{5}{3}\tan^2\theta_W M_2$,
one obtains for the lightest chargino mass the bound \cite{chargino}
$m_{\tilde\chi_1^{\pm}}>103$ GeV.
Likewise, one is also able to obtain the following 
bounds for sleptons masses \cite{sleptons}:
$m_{\tilde e}>99$ GeV,
$m_{\tilde\mu}>96$ GeV,
$m_{\tilde\tau}>87$ GeV.
Finally, we use the following 
bounds on the masses of sneutrino, the stop, the rest of squarks, and
gluinos: 
$m_{\tilde\nu}>50$ GeV,
$m_{\tilde t}>95$ GeV,
$m_{\tilde q}>150$ GeV,
$m_{\tilde g}>190$ GeV.



\item $b\to s\gamma$ 

The measurements of $B\rightarrow X_s\gamma$ decays 
at 
CLEO \cite{cleo} 
and BELLE \cite{belle},
lead to bounds on the branching ratio 
$b\to s\gamma$. In particular we impose in our computation
$2\times 10^{-4}\leq BR(b\to s\gamma)\leq 4.1\times
10^{-4}$, where the evaluation 
is carried out at the NLO accuracy in
the SM \cite{gambino}, while for the SUSY contributions only the 1-loop
diagrams have been taken into account \cite{GS}.

\item $g_{\mu}-2$

The new measurement of the anomalous
magnetic moment of the muon, $a_\mu=(g_{\mu}-2)/2$, 
in the E821 experiment at the 
BNL \cite{muon} implied apparently a 2.6$\sigma$ 
deviation from the SM predictions.
In particular, taking a 2$\sigma$ range around the E821 central value
one would have
$11\times 10^{-10}\leq a_{\mu}(E821)-a_{\mu}(SM)\leq 75\times 10^{-10}$.
However, recent theoretical computations \cite{rafael}
have shown that a significant part of this discrepancy was due to the
evaluation of the hadronic light-by-light 
contribution \cite{todos}.
As a consequence, assuming that the possible 
new physics is due to SUGRA, we impose in our computation the current
constraint
$-7\times 10^{-10}\leq a_{\mu} (SUGRA)\leq 57\times 10^{-10}$ at the
2$\sigma$ level.

\item LSP

The LSP must be an electrically neutral
(also with no strong interactions) particle,
since otherwise it 
would bind to nuclei and would be excluded as a candidate
for dark matter from unsuccessful searches for exotic heavy 
isotopes \cite{isotopes}.
As mentioned in the Introduction, the $\tilde{\chi}_1^0$ is
the LSP in most of the parameter space of SUGRA.
However, in some regions the stau can be lighter 
than the $\tilde{\chi}_1^0$. Following the above argument,
we discard these regions.

Other cosmological constraints, as e.g. the observational bounds
on the relic $\tilde{\chi}_1^0$ density,
$0.1\lsim \Omega_{\tilde{\chi}_1^0}h^2\lsim 0.3$,
will not be applied. The computation of the relic  $\tilde{\chi}_1^0$
density depends on assumptions about the evolution of the early
Universe, and therefore different cosmological scenarios give
rise to different results \cite{relic}.



\end{itemize}

\section{SUGRA predictions for the neutralino-nucleon cross section}

The cross section for the elastic scattering of  $\tilde{\chi}^0_1$
on nucleons  
has been examined exhaustively in the 
literature (for a recent re-evaluation see ref.~\cite{Ellis}).
Let us recall that
the relevant effective Lagrangian describing
the elastic scattering of  $\tilde{\chi}^0_1$ on protons and neutrons
has a spin-independent (scalar) interaction and a spin-dependent
interaction.
However, the contribution of the scalar one is generically larger and
therefore we will concentrate on it. 
This scalar interaction includes contributions from squark exchange and
neutral Higgs exchange.
Let us also remark that
the scalar cross sections for both, protons and neutrons, 
are basically equal.

In what follows we will review the possible value of the cross section 
in the framework of SUGRA. First
we will consider 
the mSUGRA scenario, where the soft terms are assumed to be universal,
and we 
will later relax this condition 
allowing non-universal soft masses.

\subsection{mSUGRA scenario}

In the mSUGRA scenario $\tilde{\chi}^0_1$ is mainly bino.
This result can be qualitatively understood from the 
well known evolution of
$m_{H_u}^2$ with the scale.
It becomes large and negative at low scales. Then 
$|\mu|$ given by eq.~(\ref{electroweak}) 
becomes also large, and in particular much larger than
$M_1$ and $M_2$. Thus,  
the lightest neutralino will be mainly gaugino, and in particular
bino, since, as mentioned in the previous section, 
at low energy $M_1=\frac{5}{3}\tan^2\theta_W M_2\approx 0.5 M_2$.

\begin{figure}[t]
\begin{center}
\epsfig{file= 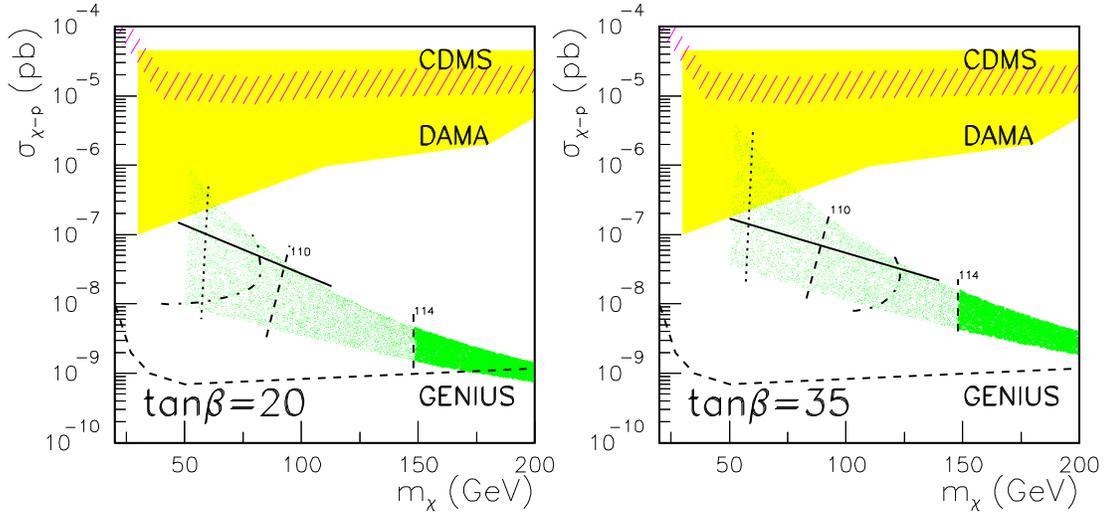,
height=7cm
}
\end{center}

\vspace{-0.5cm}

\caption{
Scatter plot of the neutralino-proton cross section
$\sigma_{\tilde{\chi}_1^0-p}$ 
as a function of 
the neutralino mass $m_{\tilde{\chi}_1^0}$ in the mSUGRA scenario, for
$\tan\beta=20$ and 35.
The dotted line corresponds to the LEP lower bound
$m_{\tilde\chi_1^{\pm}}=103$ GeV. Dashed lines correspond to different
values of the Higgs mass, 110 and 114 GeV.
The region to the left of the dot-dashed curve is excluded by
$b\to s\gamma$,
and the region above the solid line is excluded by $g_{\mu}-2$.
DAMA and CDMS current 
experimental limits and projected GENIUS limits are shown.
}
\label{ellisfig}
\end{figure}

When  $\tilde{\chi}^0_1$ is basically bino
scattering channels through Higgs ($h$, $H$) exchange are highly 
suppressed, and
as a consequence, the predicted scalar $\tilde{\chi}^0_1$-proton cross section
is well below the accessible experimental regions for low and moderate
values of $\tan\beta$
\cite{Ellis}.
Although
the cross section increases entering in the DAMA 
region \cite{experimento1,halo} when the value
of $\tan\beta$ increases \cite{Bottino,Arnowitt}, 
the experimental constraints discussed in Section~3
exclude this possibility \cite{Ellis2,Arnowitt3}.
We show this fact in
Fig.~\ref{ellisfig}. There the  $\tilde{\chi}^0_1$-proton cross section
$\sigma_{\tilde{\chi}_1^0-p}$ as a function of the neutralino mass
$m_{\tilde{\chi}_1^0}$ for
$\tan \beta=20$ and 35 is plotted.

Concerning the parameter space of the figure,
we use the following values. For the soft scalar mass
0$\leq m \leq 600$ GeV, where 
the curve associated to $m=600$ GeV 
corresponds to the minimum values of 
$\sigma_{\tilde{\chi}_{1}^{0}-p}$ in the figure.
Note that 
in order to avoid the stau being the LSP not very small values of
$m$ are in fact allowed.
The cross section is not very sensitive to 
the specific values of $A$ in a wide 
range. In particular,
we have checked that 
this is so for $\mid A/M\mid \lsim 4$. In the figure we fix
$A = M$,
corresponding to the maximum value of $\sigma_{\tilde{\chi}_{1}^{0}-p}$.
On the other hand,
the gaugino mass
$M$ is essentially fixed for a given $\tan\beta$ and 
$m_{\tilde{\chi}_1^0}$. 
We consider only positive values of $M$, since negative ones
correspond to smaller cross sections.
Finally, 
we choose to plot only the case $\mu > 0$ since for negative values of
$\mu$ the cross section is 
smaller. In addition, constraints
coming from
the $b\rightarrow s\gamma$ and $g_{\mu}-2$ processes highly reduce
the $\mu<0$ parameter space. 
Let us remark that we are using the conventions
for the SUGRA parameters entering in the Lagrangian
explained in ref.~\cite{darkcairo}.

Concerning the experimental constraints, the lower
bound in the figure $m_{\tilde{\chi}_1^0}\gsim 50$ GeV is obtained
imposing the experimental bound on the 
lightest chargino mass 
$m_{\tilde\chi_1^{\pm}}>90$ GeV. However, as mentioned in Section~3,
the most recent bound in the context of mSUGRA is now \cite{chargino}  
$m_{\tilde\chi_1^{\pm}}>103$ GeV, and we show it with
a dotted line.
We also show in Fig.~1 several values of the Higgs mass with 
dashed lines.
In addition the region to the left of the dot-dashed curve
is excluded by $b\to s\gamma$,
and the region above the solid line is excluded by $g_{\mu}-2$.

As we can see, 
the present experimental lower limit for the Higgs mass 
in mSUGRA when $\tan\beta\lsim 50$, $m_h\gsim 114.1$ GeV,
is the most important constraint, implying that only the
solid region where  $\sigma_{\tilde{\chi}_1^0-p} \lsim 10^{-8}$ pb
survives. 
Obviously, in this mSUGRA case, more sensitive detectors
producing further data are needed.
Fortunately, many dark matter detectors are being projected.
Particularly interesting is the GENIUS detector \cite{GENIUS},
where values of the cross section as low as 
$10^{-9}$ pb will be accessible, as shown in the figure.




Very large values of 
$\tan\beta$, 
like $\tan \beta \simeq 50$, 
have also been considered \cite{Mario,Ellis2,Arnowitt3}.
Although it was found, as expected, that the 
cross section is enhanced,
the 
well known
experimental limits coming from 
$b \to s \gamma$ for large $\tan\beta$, 
lead to severe constraints on the parameter space. In particular,
these constraints imply $\sigma_{\tilde{\chi}_1^0-p} \lsim 
10^{-7}$ pb.

\begin{figure}[t]
\begin{center}
\epsfig{figure=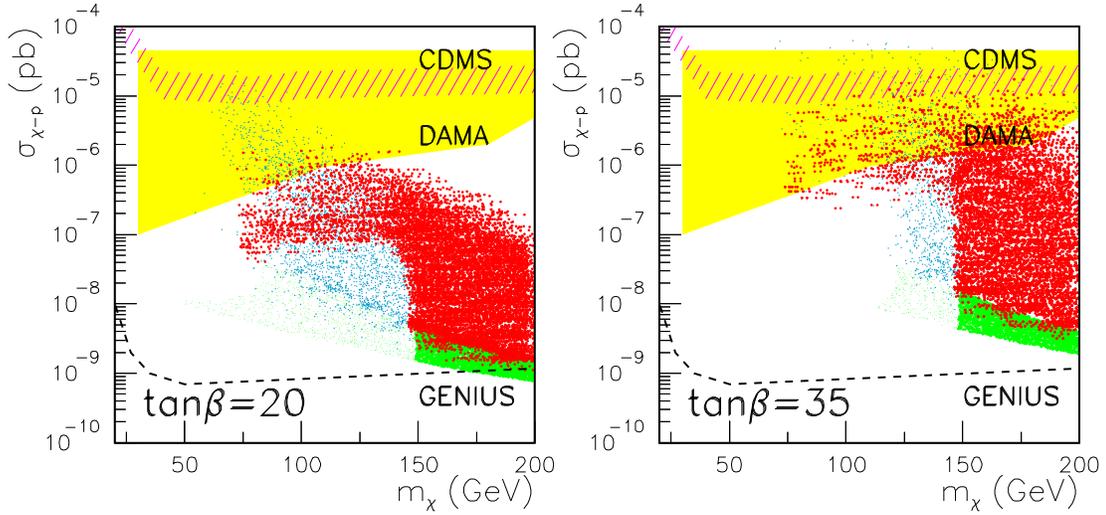, 
height=7cm}
\end{center}

\vspace{-0.5cm}

\caption{
Scatter plot of the neutralino-proton cross section
$\sigma_{\tilde{\chi}_1^0-p}$ 
as a function of 
the neutralino mass $m_{\tilde{\chi}_1^0}$ in the SUGRA scenario
with non-universal soft scalar masses, for
$\tan\beta=20$ and 35.
Only points of the parameter space fulfilling
$b \to s \gamma$ and $g_{\mu}-2$ are plotted. 
Big (red) dots correspond to points 
with $m_h\geq 114$ GeV. Small (blue) dots correspond to points
with $91 <m_h < 114$ GeV. 
The universal case (green region) 
is shown for comparison.
DAMA and CDMS current 
experimental limits and projected GENIUS limits are also shown.
}
\label{arnowit}
\end{figure}

\subsection{Scenario with non-universal soft terms}

 
It was shown recently
that non-universality allows to 
increase 
the neutralino-proton cross section. 
The key point
consists of reducing the value of $|\mu|$.
Following the discussion in the previous subsection, 
the smaller $|\mu|$ is,
the larger the Higgsino components of the lightest neutralino become.
Eventually, $|\mu|$ will be of the order of $M_1$, $M_2$
and $\tilde{\chi}_1^0$ will be a mixed Higgsino-gaugino state.
Indeed scattering channels through Higgs exchange 
are now important and their contributions to the cross section
will increase it.
This can be carried out with non-universal 
scalar masses \cite{Bottino,Arnowitt,Arnowitt3}
and/or gaugino masses \cite{Nath2,darkcairo}.
We can have
a qualitative
understanding 
from the following.
When $m_{H_u}^2$($m_{H_d}^2$) at $M_{GUT}$ increases(decreases),
its negative(positive) contribution at low energy in eq.~(\ref{electroweak})
is less important. Likewise, when $m_{Q_{L}}^2$ and $m_{u_{R}}^2$
at $M_{GUT}$ decrease, due to their contribution proportional
to the top Yukawa coupling in the RGE of $m_{H_u}^2$ the negative
contribution of the latter to $\mu^2$ is again less important.
%
%

This effect can be seen in Fig.~\ref{arnowit}, where
a scatter plot of
$\sigma_{\tilde{\chi}_{1}^{0}-p}$ as a function of $m_{\tilde{\chi}_{1}^{0}}$
for a scanning of the soft scalar masses is shown.
This scanning is explained in detail e.g. in ref.~\cite{darkcairo}.
As in Fig.~\ref{ellisfig}, we are taking $A=M$.
For comparison we superimpose also the region (green area) 
obtained in Fig.~\ref{ellisfig} 
with universality.
We see that non-universal scalar masses help in increasing the value of
$\sigma_{\tilde{\chi}_{1}^{0}-p}$.
In the figure we only plot the points of the parameter space fulfilling
$b \to s \gamma$ and $g_{\mu}-2$. In particular,
the big (red) dots correspond to points 
with $m_h\geq 114$ GeV. As a conservative approach,
we also plot with small (blue) dots those points
with $91 <m_h < 114$ GeV (as discussed in Section~3,
this possibility may arise when the mSUGRA scenario is relaxed).
For $\tan\beta=20$, both kind of points enter in the DAMA region.
In particular, those with $m_h < 114$ GeV have an upper bound for
the cross section 
$\approx 10^{-5}$ pb, whereas
for those with $m_h > 114$ GeV the upper bound is
$\approx 10^{-6}$ pb.
For $\tan\beta=35$, the area of points entering in the DAMA region
increases a lot, and now, even points with $m_h > 114$ GeV
can have $\sigma_{\tilde{\chi}_{1}^{0}-p} \approx 10^{-5}$ pb.
Let us finally remark that the 
non-universality in the Higgs sector gives the most important
effect, and the non-universality in the sfermion sector only increases
slightly the cross section.

\begin{figure}[t]
\begin{center}
\epsfig{figure=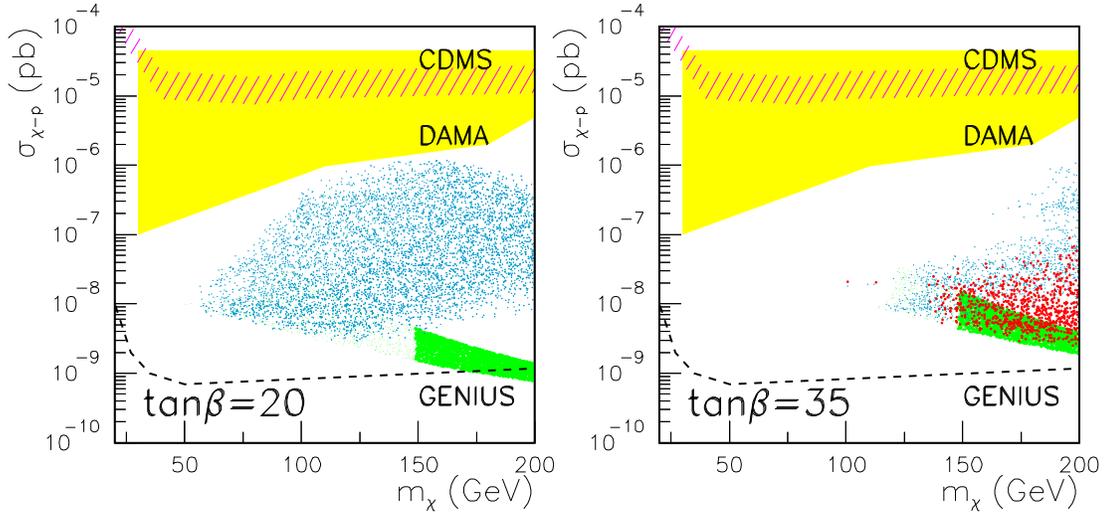,
height=7cm}
\end{center}

\vspace{-0.5cm}

\caption{
The same as in Fig.~\ref{arnowit}
but using non-universal soft 
gaugino masses. 
\label{nunivg.eps}}
\end{figure}

On the other hand, concerning gaugino masses,
it is worth noticing that
$M_3$ appears in the RGEs of squark masses, so e.g.
their contribution proportional
to the top Yukawa coupling in the RGE of $m_{H_u}^2$ will
do this less negative if $M_3$ is small, and therefore $\mu^2$ will become 
smaller in this case.
Taking into account this effect, 
we show in Fig.~\ref{nunivg.eps},
for $\tan\beta=20$ and 35,
a scatter plot of
$\sigma_{\tilde{\chi}_{1}^{0}-p}$ as a function of $m_{\tilde{\chi}_{1}^{0}}$
for the scanning of the parameters $M_{3}$, $M_{2}$, $M_{1}$  explained
in ref.~\cite{darkcairo}.
Although points with a large cross section entering in the DAMA region
do exist \cite{darkcairo}, they are, at the end of the day, forbidden 
when 
$b \to s \gamma$ and $g_{\mu}-2$ constraints are imposed.

Clearly, in this sense, SUGRA scenarios with non-universal scalars
are favored with respect to scenarios with non-universal gauginos.

\begin{figure}[t]
\begin{center}
\epsfig{file= 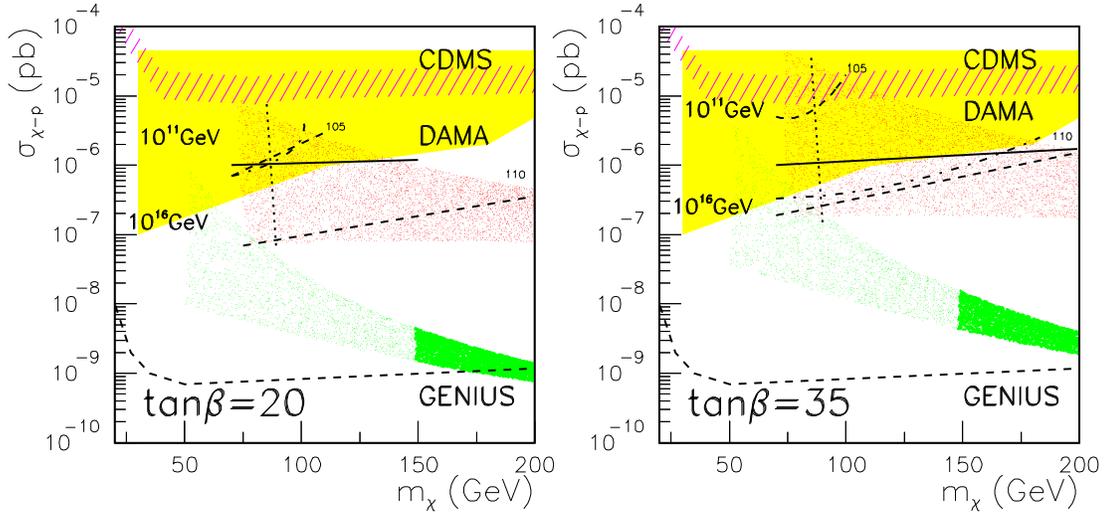, 
height=7cm}
\end{center}
\vspace{-0.5cm}
\caption{
The same as in Fig.~\ref{ellisfig}
but including the case 
with intermediate scale,
$M_I=10^{11}$ GeV.
Now the Higgs masses 105 and 110 GeV are shown. 
}
\label{sigmaM_I}
\end{figure}

\section{Scenario with intermediate unification scale}

The analyses of 
$\sigma_{\tilde{\chi}_1^0-p}$ 
in SUGRA, described in the previous section,
were performed assuming the
unification scale
$M_{GUT} \approx 10^{16}$ GeV.
However, there are several interesting arguments in favor of SUGRA scenarios
with scales 
$M_I\approx 10^{10-14}$ GeV.
These can be found e.g. in ref.~\cite{darkcairo} and references therein.
Thus to use the value of the initial scale,
say $M_I$, as a free parameter for the running of the soft terms
is particularly interesting.
In this sense, it was recently 
pointed out \cite{muas} 
that 
$\sigma_{\tilde{\chi}_1^0-p}$
is very sensitive to the variation of the initial scale 
for the running of the soft terms.
For instance, by taking $M_I=10^{10-12}$ GeV rather than 
$M_{GUT}$, regions in the parameter space of mSUGRA can be  
found 
where  $\sigma_{\tilde{\chi}_1^0-p}$ is two orders of magnitude
larger than for $M_{GUT}$.


The fact that smaller initial scales imply a larger 
$\sigma_{\tilde{\chi}_1^0-p}$ 
can be understood from the variation in the value of 
$\mu$ with $M_I$.
One observes that, for $\tan\beta$ fixed, the smaller the initial
scale for the running is, the smaller the numerator in the
first piece of eq.(\ref{electroweak}) becomes. 
This can be understood qualitatively from 
the well known evolution of $m_{H_d}^2$ and $m_{H_u}^2$ with the
scale.
Clearly, the smaller the initial scale is, the shorter the
running becomes. As a consequence, 
also the less important the positive(negative) contribution 
$m_{H_d}^2$($m_{H_u}^2$) to $\mu^2$ in eq.(\ref{electroweak}) becomes.
Thus 
$|\mu|$ decreases, and therefore, as discussed in the previous section,
$\sigma_{\tilde{\chi}_1^0-p}$ increases.
This is shown in 
Fig.~\ref{sigmaM_I}, where the result
for the scale $M_I=M_{GUT}$ is compared 
with the result for 
the intermediate scale $M_I=10^{11}$ GeV.

Clearly, in the latter scenario the Higgs mass decreases
and all points would be excluded if we impose the mSUGRA bound
$m_h>114$ GeV.
On the other hand, taking into account the conservative approach
discussed in Section~3, since all points in the figure have $m_h>100$ GeV
we will not discard them.
Then, the most important constraint for $\tan\beta=20$
is due to $g_{\mu}-2$, excluding a large area of points in the
DAMA region. 
We can understand this from the results of ref.~\cite{muonmio}.
There, $a_{\mu}$ versus $m_{\tilde{\chi}_1^0}$ is plotted, and one
can observe that for $m_{\tilde{\chi}_1^0}\leq 150$ GeV and $m$ small 
too large values of $a_{\mu}$, beyond
the experimental upper bound, are obtained. 
Note that the area left in DAMA by this constraint
and the mSUGRA one
$m_{\tilde\chi_1^{\pm}}>103$ GeV
may be increased if we relaxed the latter bound to e.g. 90 GeV.
We see in Fig.~\ref{sigmaM_I}
that the range 75 GeV $\lsim m_{\tilde{\chi}_1^0}\lsim$ 125 GeV is now
consistent 
with DAMA limits. 
However, for $\tan\beta=35$ all points in the DAMA region are
excluded because of $b \to s \gamma$.

\begin{figure}[t]
\begin{center}
\epsfig{file= 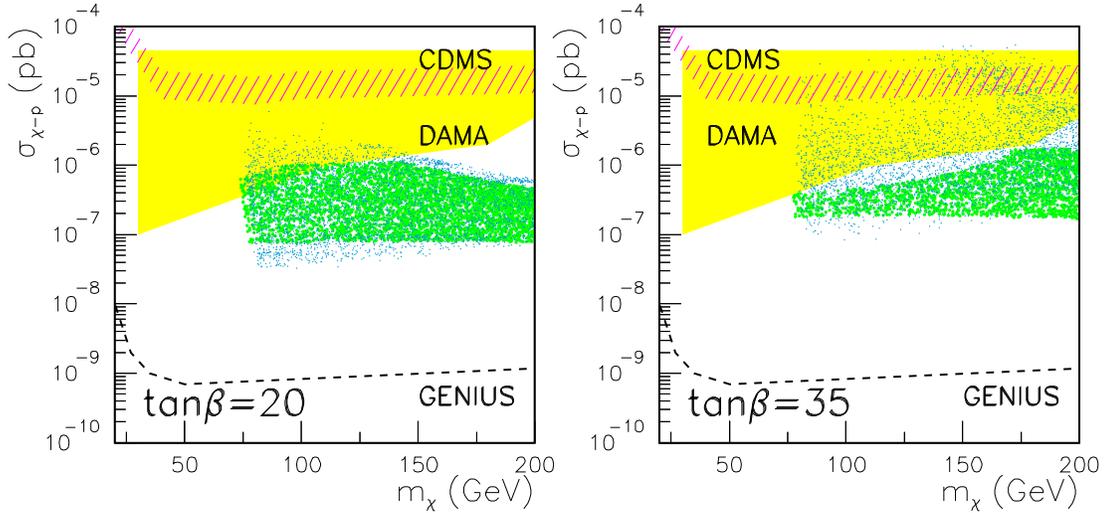, 
height=7cm}
\end{center}
\vspace{-0.5cm}
\caption{
Scatter plot of the neutralino-proton cross section
$\sigma_{\tilde{\chi}_1^0-p}$ 
as a function of 
the neutralino mass $m_{\tilde{\chi}_1^0}$ in the SUGRA scenario
with the intermediate scale
$M_I=10^{11}$ GeV and 
non-universal soft scalar masses, for
$\tan\beta=20$ and 35.
Only points of the parameter space fulfilling
$b \to s \gamma$ and $g_{\mu}-2$ are plotted. 
(Blue) dots correspond to points
with $91 <m_h < 114$ GeV. 
The universal case (green region) 
is shown for comparison.
DAMA and CDMS current 
experimental limits and projected GENIUS limits are also shown.
}
\label{sigmaM_I2}
\end{figure}

Given the above situation concerning the enhancement of the 
neutralino-proton cross section through non-universal scalars
and intermediate scales, it is worth analyzing the
combination of both possibilities.
This is shown in Fig.~\ref{sigmaM_I2}.
Although no points with $m_h > 114$ GeV enter in the DAMA region,
many points with 
$91 <m_h < 114$ GeV do enter and 
are allowed by all the other constraints. For
$\tan\beta=35$ the allowed area is extremely large and
we can even have points entering in the region
excluded by CDMS.

\section{Superstring scenario with D-branes}

In the previous section the analyses were performed assuming intermediate
unification scales.
In fact, this situation can be inspired by superstring theories, since
it was recently 
realized that 
the string scale may be anywhere between the weak and the Plank 
scale.
For example, embedding the SM inside D3-branes in type I
strings, the string scale is given by
$M_I^4= \alpha M_{Planck} M_c^3/{\sqrt 2}$,
where $\alpha$ is the gauge coupling and $M_c$ is the compactification scale. 
Thus one gets $M_I\approx 10^{10-12}$ GeV with $M_c\approx 10^{8-10}$ GeV.
In addition,
D-brane constructions are explicit scenarios where not only
intermediate scales arise naturally but also 
non-universality.

\begin{figure}[t]
\begin{center}
\epsfig{file= 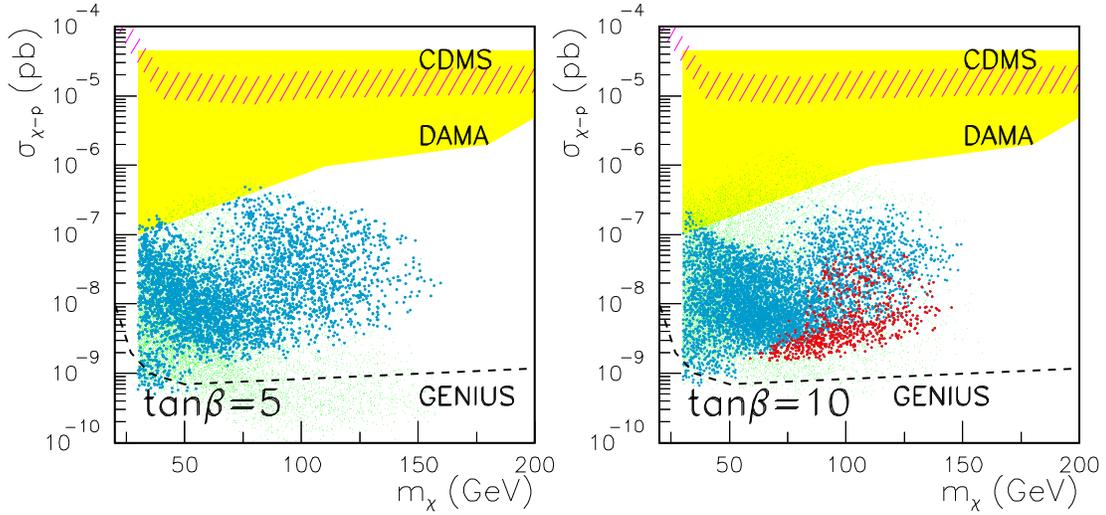, 
height=7cm
}
\end{center}
\vspace{-0.5cm}
\caption{
Scatter plot of the neutralino-proton cross section
$\sigma_{\tilde{\chi}_1^0-p}$ 
as a function of 
the neutralino mass $m_{\tilde{\chi}_1^0}$ in the D-brane scenario
with 
the string scale $M_I=10^{12}$ GeV
discussed
in the text, and for $\tan\beta= 10$ and 15.
Only the big (red and blue) dots 
fulfil $b \to s \gamma$ and $g_{\mu}-2$ constraints.
The red ones 
correspond to points 
with $m_h\geq 114$ GeV whereas the blue ones correspond
to points
with $91 <m_h < 114$ GeV. 
DAMA and CDMS current 
experimental limits and projected GENIUS limits are also shown.
}
\label{dbrane}
\end{figure}

The crucial point in these analyses 
is the D-brane origin of the $U(1)_Y$ gauge group
as a combination of other $U(1)$'s, 
and its influence on the matter distribution in these scenarios.
In particular, 
scenarios with the gauge group and particle content of the
SUSY SM lead naturally to intermediate values for the
string
scale, in order to reproduce the value of gauge couplings
deduced from experiments. In addition, the soft terms 
turn out to be generically non-universal.
Due to these results, 
large cross sections
can be obtained \cite{nosotros}.

Let us consider for example a type I string scenario 
where the gauge group 
$U(3)\times U(2)\times U(1)$, giving rise to
$SU(3)\times SU(2)\times U(1)^3$, arises
from three different types of D-branes, 
and therefore the gauge couplings are non-universal. 
Here
$U(1)_Y$ is a linear combination
of the three $U(1)$ gauge groups arising from $U(3)$, $U(2)$ and
$U(1)$ within the three different D-branes. 
As shown in ref.~\cite{nosotros},  
this leads
to the string scale $M_I = 10^{12}$ GeV. 
Likewise,
assuming 
that only the
dilaton ($S$) and moduli ($T_i$) fields contribute to SUSY
breaking, it was found  that the soft terms  are generically
non-universal.

Fig.~\ref{dbrane} displays a scatter plot of 
$\sigma_{\tilde\chi_1^0-p}$ as a function of the
neutralino mass $m_{\tilde\chi_1^0}$ for a scanning of the parameter
space
described in ref.~\cite{nosotros}. In particular,
the gravitino mass $m_{3/2}\leq 300$ GeV is taken, since
larger values will always produce a cross section below DAMA limits.
The cases $\tan\beta=10$ and 15 are shown in Fig.~\ref{dbrane}.
Although regions of the parameter
space consistent with DAMA limits exist, the 
$b \to s \gamma$ and $g_{\mu}-2$ constraints forbid 
most of them. The latter are shown with small (green) points,
and they have $91 <m_h < 114$ GeV. For example, 
to understand this result for $g_{\mu}-2$ is simple.
In ref.~\cite{muonmio} $a_{\mu}$ versus $m_{3/2}$ is plotted, and one
can observe that for $m_{3/2}\lsim 300$ GeV 
too large values of $a_{\mu}$, beyond
the experimental upper bound, are obtained. 
In Fig.~\ref{dbrane} only regions with big (red and blue) dots 
fulfil the above mentioned constraints.
The red ones
correspond to points 
with $m_h\geq 114$ GeV whereas the blue ones correspond
to points
with $91 <m_h < 114$ GeV. 
It is worth noticing that the larger $\tan\beta$ is, 
the smaller the regions allowed by the experimental constraints become.
For example, increasing
$\tan\beta$ the value of 
$a_{\mu}$ turns out to be larger and may exceed the experimental bounds.

\section{Conclusions}

We have reviewed the direct detection of 
supersymmetric dark matter in the light of recent experimental results.
They require a large cross section of the order of
$10^{-6}$ pb. 
Although in the context of the mSUGRA scenario one cannot 
obtain this value once the experimental constraints are imposed,
this is possible for other scenarios.
This is the case of SUGRA models with non-universal soft scalar masses.
Likewise, scenarios with intermediate unification scale
may also produce this value if some experimental bounds are relaxed.
\\

\noindent {\bf Acknowledgements}

\noindent
We thank S. Khalil and E. Torrente-Lujan
as co-authors of some of the works reported in
this paper. D.G. Cerde\~no acknowledges the financial support
of the Comunidad de Madrid through a FPI grant.
E. Gabrielli is grateful to the CERN Theory Division and
the Department of Theoretical Physics of Madrid 
Aut\'onoma University, where 
part of this work has been done, for their warm hospitality.
The work of C. Mu\~noz was supported in part by the Ministerio de 
Ciencia y Tecnolog\'{\i}a under contract FPA2000-0980, 
and the European Union under contract 
HPRN-CT-2000-00148.

\end{document}